\documentclass[10pt,conference]{IEEEtran}

\IEEEoverridecommandlockouts
\usepackage{cite}
\usepackage{amsmath,amssymb,amsfonts}
\usepackage{url}
\usepackage{booktabs}
\usepackage{xurl}

\usepackage{multirow}
\usepackage{multicol}
\usepackage{algorithmic}
\usepackage{graphicx}
\usepackage{textcomp}
\usepackage{xcolor}
\usepackage{color}
\usepackage{listings}
\usepackage{xspace}
\usepackage{booktabs}
\usepackage{array}
\usepackage{enumitem}
\usepackage{tcolorbox}

\usepackage[colorlinks]{hyperref} 

\definecolor[named]{ACMBlue}{cmyk}{1,0.1,0,0.1} 
\definecolor[named]{ACMYellow}{cmyk}{0,0.16,1,0} 
\definecolor[named]{ACMOrange}{cmyk}{0,0.42,1,0.01} 
\definecolor[named]{ACMRed}{cmyk}{0,0.90,0.86,0} 
\definecolor[named]{ACMLightBlue}{cmyk}{0.49,0.01,0,0} 
\definecolor[named]{ACMGreen}{cmyk}{0.20,0,1,0.19} 
\definecolor[named]{ACMPurple}{cmyk}{0.55,1,0,0.15} 
\definecolor[named]{ACMDarkBlue}{cmyk}{1,0.58,0,0.21} 

\usepackage{inconsolata}
\usepackage{mathptmx} 

\newcommand{\data}{MEIC\xspace}
\usepackage{bbm}
\def\BibTeX{{\rm B\kern-.05em{\sc i\kern-.025em b}\kern-.08em
    T\kern-.1667em\lower.7ex\hbox{E}\kern-.125emX}}
\begin{document}
\hypersetup{
linkcolor=ACMPurple, citecolor=ACMPurple, urlcolor=ACMDarkBlue, filecolor=ACMDarkBlue } 

\newcommand{\wcz}[1]{\textcolor{violet}{{#1}}}
\newcommand{\zj}[1]{\textcolor{red}{{#1}}}

\newcommand{\finding}[2]{
\begin{tcolorbox}[width=\linewidth,boxrule=0pt,top=1pt, bottom=1pt, left=1pt,right=1pt, colback=gray!20,colframe=gray!20]
\textbf{Finding #1:} 
{#2}
\end{tcolorbox}}
\newcommand{\yun}[1]{\textcolor{blue}{{#1}}}

\title{Exploring Multi-Lingual Bias of Large Code Models in Code Generation
}

\author{
    \IEEEauthorblockN{Chaozheng Wang\IEEEauthorrefmark{2}, Zongjie Li\IEEEauthorrefmark{3}, Cuiyun Gao\IEEEauthorrefmark{2}\IEEEauthorrefmark{1}\thanks{* Cuiyun Gao is the corresponding author.}, Wenxuan Wang\IEEEauthorrefmark{2}\\ Ting Peng\IEEEauthorrefmark{4}, Hailiang Huang\IEEEauthorrefmark{4}, Yuetang Deng\IEEEauthorrefmark{4}, Shuai Wang\IEEEauthorrefmark{3}, Michael R. Lyu\IEEEauthorrefmark{2}}
    \IEEEauthorblockA{\IEEEauthorrefmark{2} The Chinese University of Hong Kong, Hong Kong, China}
    \IEEEauthorblockA{\IEEEauthorrefmark{3} Hong Kong University of Science and Technology, Hong Kong, China}
    \IEEEauthorblockA{\IEEEauthorrefmark{4} Tencent Inc., Guangzhou, China}
    \IEEEauthorblockA{\{czwang23, wxwang, lyu\}@cse.cuhk.edu.hk,\{zligo, shuaiw\}@cse.ust.hk, cuiyungao@outlook.com, }
}

\maketitle

\begin{abstract}
Code generation aims to synthesize code and fulfill functional requirements based on natural language (NL) specifications, which can greatly improve development efficiency. In the era of large language models (LLMs), large code models (LCMs) have been recently proposed to generate source code. LCMs can generate highly feasible solutions for programming problems described in natural language. Despite the effectiveness, we observe a noticeable multilingual bias in the generation performance of LCMs. Specifically, LCMs demonstrate proficiency in generating solutions when provided with instructions in English, yet may falter when faced with semantically equivalent instructions in other NLs such as Chinese. Moreover, the ability of LCMs to generate code exhibits variety across different programming languages (PLs), such as Python and C++. The observed phenomenon indicates the presence of multi-lingual bias within the generative capabilities of LCMs, which has remained unexplored.


In this paper, we aim to investigate the multi-lingual bias that exists in current LCMs. First, we initiate our investigation by constructing the first multi-lingual evaluation benchmark \textit{X-HumanEval-X}, enabling us to systematically evaluate the extent of multi-lingual bias that exists in current LCMs. In our large-scale experiments on nine popular LCMs, we observe a pronounced multi-lingual bias of LCMs in code generation, including multi-NL and multi-PL bias. Specifically, when using Chinese instructions, the code generation capabilities of LCMs decrease by at least 13\% in terms of the Pass@1 metric. Furthermore, LCMs perform variously across different programming languages, e.g., the performance gap between Python and C++ reaches as high as 20.9\%. Then we explore the bias in the prompting phase and find that prompting LCMs through one-step and multi-step translation aids in mitigating such bias. We further explore the impact of instruction tuning based on a self-constructed multi-lingual dataset Multi-EvolInstruct-Code (\data) that contains two natural languages (i.e., English and Chinese) and more than twenty programming languages. Experiments on the nine popular LCMs demonstrate that the instruction tuning substantially reduces the multi-lingual bias (e.g., decreasing the multi-NL bias and multi-PL bias by up to 84\% and 40\%, respectively), while enhancing the efficacy of LCMs in code generation (e.g., increasing the Pass@1 metric by 31\%$\sim$46\%). We finally provide insights and implications for researchers and developers aimed at mitigating the multi-lingual bias and improving the code generation capabilities of LCMs.
\end{abstract}

\begin{IEEEkeywords}
code generation, multi-lingual, large language models
\end{IEEEkeywords}

\section{Introduction}
Code generation is a fundamental task within the domain of code intelligence, designed to interpret natural language instructions and produce corresponding code snippets that fulfill specifications. 
With the development of large language models (LLMs), the field of code intelligence has witnessed the emergence of large code models (LCMs) that are specifically tailored for programming tasks including code generation~\cite{guo2024deepseek, chen2021evaluating, yang2024unveiling, ali2024memory, roziere2023code}.
These models, trained on extensive datasets of source code, have substantially enhanced the efficiency of code generation, thereby considerately reducing the coding burden
for developers. The remarkable capabilities of LCMs have garnered attention from users globally, spanning diverse roles within the software development process. This diversity highlights the wide range of purposes and ways in which users engage with LCMs, including the use of multiple natural languages (NLs) and the expectation for LCMs to produce code snippets across different programming languages (PLs).


\begin{figure*}[t]
    \centering
    \includegraphics[width=0.92\textwidth]{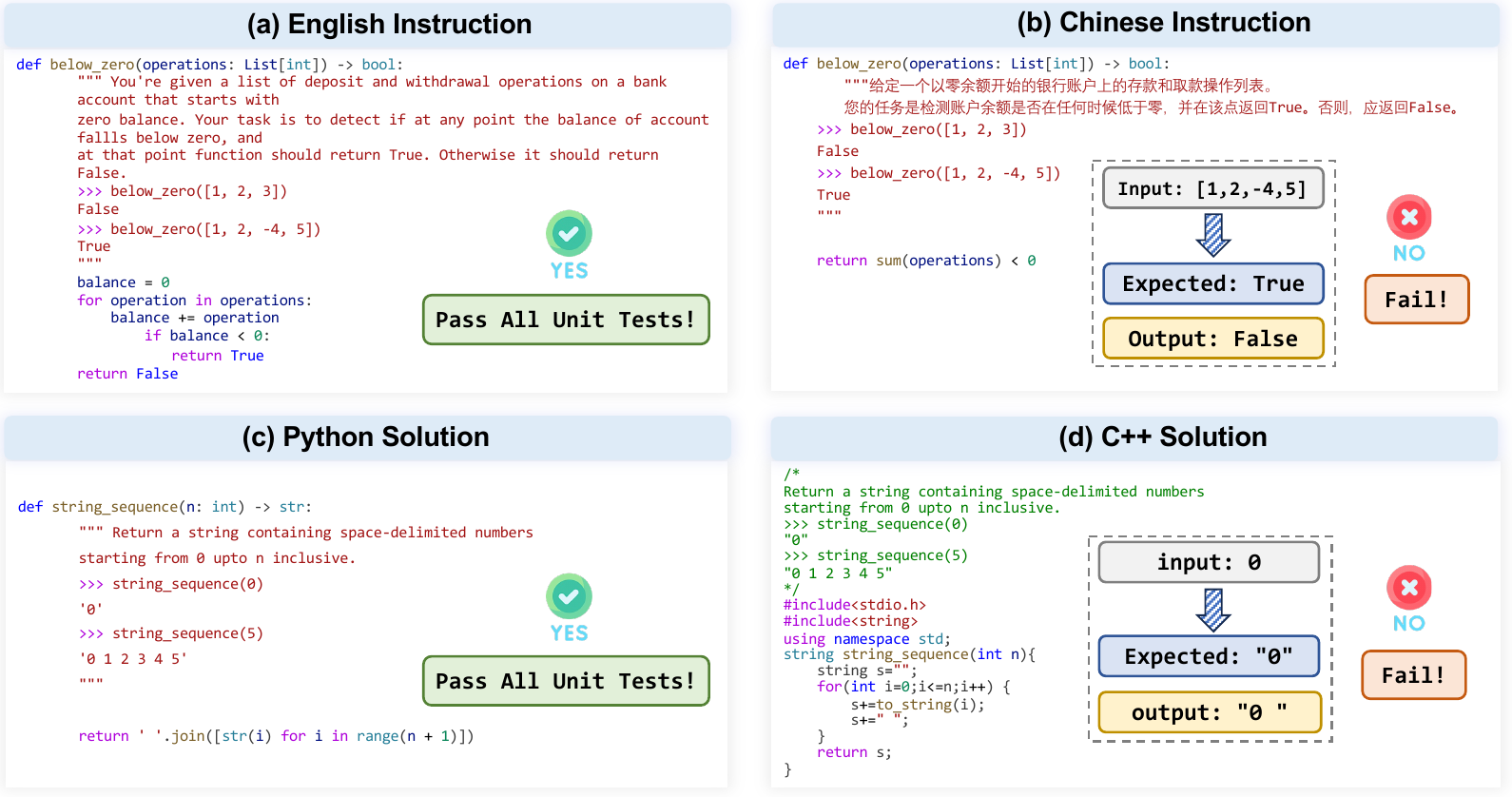}
    \caption{Motivation example of the bias that exists among multiple natural languages ((a) and (b)) and programming languages ((c) and (d)). The experimented LCM is DeepSeek-Coder 33B.}
    \label{fig:example}
    \vspace{-12pt}
\end{figure*}

In the domain of code generation, the broad user base of LCMs introduces a great demand for multi-lingual competencies, encompassing both multi-NL and multi-PL support.
From the perspective of multi-NL, users expect these models to understand and generate content across multiple languages. According to the report~\cite{chatgptuser}, a large population of users (around 80\%) are from non-English speaking regions, including but not limited to Japan and China, thereby underscoring the global demand for multi-lingual support.
From the multi-PL perspective, the demand for multi-PL in LCMs arises from the distinct design principles and exclusive features of each language, which have contributed to their popularity within specific communities and development \cite{plstatis}. 
For instance, Copilot \cite{copilot}, an LCM-based intelligence programming assistant, receives and processes user requests to generate source code in a variety of programming languages, including Python, C++, and JavaScript \cite{copilotlan}.
Despite the remarkable demands of users in using LLMs to understand and generate different languages, they are still faced with multi-lingual bias that undermines their universal effectiveness~\cite{Lai2023ChatGPTBE, Bang2023AMM, Shi2022LanguageMA, Wang2023AllLM, Peng2023TowardsMT}. 
Existing research~\cite{Lai2023ChatGPTBE} highlights that LLMs tend to exhibit superior performance in English compared to other languages, a discrepancy attributed primarily to the predominance of English in the training datasets.
Similarly, such a multi-lingual bias is also observed in current LCMs.
As the example shown in Figure \ref{fig:example} (a) and (b), DeepSeek-Coder 33B \cite{guo2024deepseek} generates the correct solutions in response to the English instructions but fails with the semantically equivalent Chinese instructions. Additionally, the performance of LCMs varies across different programming languages. For instance, when given the same English instruction, DeepSeek-Coder 33B effectively addresses the task in Python (c) but fails to account for a crucial condition when generating code in C++ (d), demonstrating the model's bias in handling various programming languages.
However, the multi-lingual bias in the domain of code generation has remained under-explored, which necessitates a comprehensive investigation on systematically assessing and mitigating such bias.

In this paper, we aim to investigate the multi-lingual bias of LCMs in the code generation task. To facilitate this investigation, we first construct a benchmark \textit{X-HumanEval-X}, which contains instructions of two NLs (i.e., English and Chinese due to their popularity \cite{nlstatis}) as well as corresponding solutions in three PLs (including Python, Java, and C++).
Based on the \textit{X-HumanEval-X},
we evaluate the code generation performance of nine popular LCMs across different NLs and PLs, aiming to analyze
the multi-lingual bias inherent in LCMs. Then we proceed to investigate the potential for mitigating this bias via training-free methods (e.g., in the prompting phase) and training-based methods. Specifically, for training-free methods, we explore mitigating the bias by translating the Chinese instructions into English and analyze three prompting strategies including one-step translation, multi-step translation, and self-translation.
For training-based methods, we explore the mitigation of bias during the training phase, specifically focusing on the instruction tuning phase. Instruction tuning, also called supervised fine-tuning, is a pivotal step in adapting language models, aiming at refining
LCMs with high-quality examples of instructions paired with their corresponding answers 
\cite{zhang2023instruction}.
Therefore,
we construct a multi-lingual dataset named Multi-EvolInstruct-Code (\data) based on EvolInstruct \cite{luo2023wizardcoder}. 
\data comprises 91,766 training instances related to code generation, encompassing two natural languages (i.e., English and Chinese) and over twenty programming languages. 
By instruction-tuning the LCMs with \data, we explore how the choice of languages and training methods affect the multi-lingual bias of LCMs in the code generation task.

Through extensive experiments on nine popular LCMs including StarCoder~\cite{li2023starcoder}, CodeLlama \cite{codellama}, and DeepSeek-Coder \cite{guo2024deepseek}, we achieve the following findings. 

1) \textbf{Current LCMs exhibit
substantial multi-lingual bias in code generation.} We uncover two dimensions of the multi-lingual bias within current LCMs including (a) \textbf{Multi-natural language bias.}
When presented with instructions in English and Chinese that convey the same intent, we observe that the average Pass@1 rate on the \textit{X-HumanEval-X} dataset experiences a minimum decline of 13\% when switching from English to Chinese instructions.
This highlights the presence of bias in the multi-natural language understanding capabilities of LCMs. (b) \textbf{Multi-programming language bias.} Given the same instruction, LCMs may successfully generate accurate solutions in widely-used programming languages such as Python and Java but struggle with 
more intricate
languages, such as C++. Specifically, the average code generation performance gap across different PLs reaches as high as 23.7\% in terms of the average Pass@1, thereby substantiating the existence of the notable multi-programming language bias.

2) \textbf{Prompting LCMs through translation mitigates the multi-NL bias}. In addition, our exploration of instruction translation reveals that translation-based prompting strategies can mitigate multi-NL bias. Specifically, for one-step and multi-step translation strategies that employ
third-party translation tools (e.g., Google Translation) to convert instructions from Chinese into English diminish the multi-NL bias, reducing the bias from 17.2\% to as low as 3.8\% in terms of the averaged Pass@1. However, we observe that the code generation performance based on the self-translation prompting experiences a drastic decrease at the ratio of 62.3\%, even enlarging the multi-NL bias.

3) \textbf{Instruction tuning aids in mitigating the multi-lingual bias while enhancing the performance of LCMs in code generation}. 
Experiments on the nine popular LCMs demonstrate that the instruction tuning substantially reduces the multi-lingual bias (e.g., decreasing the multi-NL bias and multi-PL bias by up to 84\% and 40\%, respectively), and enhances the efficacy of LCMs in code generation (e.g., increasing the Pass@1 metric by 31\%$\sim$46\%). 
The results indicate that increasing the diversity of both NLs and PLs in the training data is beneficial for boosting the overall performance of LCMs in code generation and reducing the multi-lingual bias at the same time. 

In summary, this paper makes the following contributions:
\begin{itemize}
    \item To the best of our knowledge, we are the first to dive into the multi-lingual bias of LCMs from both the natural language and programming language perspectives in the code generation task.
    \item We construct the first multi-lingual benchmark \textit{X-HumanEval-X} and one instruction tuning dataset \data that contains two natural languages and more than twenty programming languages. 
    \item We conduct extensive experiments on nine popular LCMs and provide findings for researchers and developers that aim to mitigate the multi-lingual bias and improve the code generation capabilities of LCMs. 
\end{itemize}

\section{Overview}\label{sec:moti}
\subsection{Research Questions}
In this paper, we mainly investigate the following research questions through
experiments.

\begin{enumerate}[label=\bfseries RQ\arabic*:,leftmargin=.5in]
    \item To what extent does the multi-lingual bias exist in LCMs in the code generation task?
    \item Whether the multi-lingual bias can be mitigated in the prompting phase?
    \item Whether the multi-lingual bias can be mitigated in the instruction tuning phase? Specifically, we explore two sub-questions,
    including 1) \textbf{RQ3.1:} How do different natural languages used in the instruction tuning phase affect the LCMs' multi-NL understanding performance?
     2) \textbf{RQ3.2:} How do different programming languages affect LCMs' multi-PL generation performance?
    
\end{enumerate}

In the following part of this section, we introduce the overview of our exploration in details.


\subsection{Benchmark Construction}\label{sec:benchmark}
To initiate our investigation and answer the first research question, we construct the first multi-lingual evaluation benchmark \textit{X-HumnEval-X} and study the multi-lingual bias from two aspects including multi-NL understanding bias and multi-PL generation bias, respectively.

For assessing multi-programming language (multi-PL) generation bias, we select three representative PLs including Python, Java, and C++ by utilizing the HumanEval-X benchmark \cite{zheng2023codegeex}. HumanEval-X serves as a variation of the original HumanEval benchmark \cite{chen2021evaluating}, specifically adapted to include not just Python but also other PLs. This choice is motivated by the widespread use and popularity of these programming languages \cite{plstatis}. Our goal is to explore and quantify the performance gaps when generating solutions across varied programming languages.

For the bias in multi-natural language understanding, we choose to employ ChatGPT \cite{chatgpt} (GPT-3.5-turbo version) to translate the benchmark's original English instructions into Chinese due to its promising translation performance \cite{jiao2023chatgpt}. This procedure is taken to construct a Chinese version of the benchmark, given that the initial set of instructions was exclusively in English. To ensure precision, the first two authors independently review every translated result and make corrections when necessary. Then the two authors discuss the corrections and reach a consensus, confirming the accuracy of the corrected version. This ensures that the instructions are correctly represented and can be effectively used for evaluating the performance of the models with Chinese instructions.

\subsection{Exploration on Mitigating Multi-Lingual Bias in Prompting}
Given the observed bias in the understanding capabilities of LCMs, a straightforward approach to mitigate this bias involves translating Chinese prompts into English prior to prompting the LCMs. Thus, to answer RQ2,  which explores the potential for reducing multilingual bias during the prompting phase, we preceding experiments that involve translating Chinese instructions into English. Specifically, we explore the following methods:
1) Employing the LCMs themselves for self-translating the Chinese instructions \cite{Shi2022LanguageMA, hu2020xtreme}, denoted as \textbf{Self-Translation}, and 2) Utilizing external translation tools, such as Google Translation \cite{googletran}, to convert Chinese instructions into English. Moreover, we apply translation tools in two manners which include
\textbf{One-Step Translation} and \textbf{Multi-Step Translation}. One-Step Translation utilizes translation tools to directly translate all Chinese instructions into English, including the example cases. For Multi-Step Translation,  individual statements are translated into English separately, aiming for potentially greater accuracy or context preservation in each translation step.

\subsection{Mitigating Multi-Lingual Bias in Instruction Tuning}
In RQ3, we explore mitigating the multi-lingual bias of LCMs in the instruction tuning phase.  Specifically, we delve into how the choices of the data and training methods of instruction tuning affect
the multi-lingual bias, with the exploration details as below.

\subsubsection{Instruction Tuning Dataset Construction}
To study the multi-lingual bias from both natural language and programming language perspectives, we first construct an instruction tuning dataset that contains code generation instructions and their corresponding output answers named \data. Specifically, we utilize the Evol-Instruct \cite{xu2023wizardlm, luo2023wizardcoder} technique to construct the dataset. Following previous work \cite{luo2023wizardcoder}, we start from an open-source dataset Code-Alpaca \cite{codealpaca} that contains 20K instruction-output pairs as seed instructions, and extend their depth and width via ChatGPT \cite{chatgpt}. The details of how we extend instructions are shown in our anonymous repository.
For the extended instructions, we also feed them into GPT4 (with the API version of GPT-4-1106) to obtain output answers. 
Given that the generated dataset is purely in English, we remark the dataset as $D_{Eng}$. In summary, the dataset contains 91,766 training instances (i.e., an input instruction and output answer pair) in more than 20 programming languages such as Python, JavaScript, and SQL.

After getting the instruction tuning dataset through Evol-Instruct, we also translate the dataset into Chinese via ChatGPT,
which we remark as $D_{Chi}$. The first two authors also randomly sample 1,000 instances of the dataset, achieving a 99\% confidence level with a confidence interval of 0.88\%. Upon reviewing the translation results, we concur that 988 instances are accurately translated. Hence, it can be asserted that the dataset, translated using ChatGPT, maintains an accuracy rate of 98\%, indicating the quality of ChatGPT's translation. Totally, $D_{Eng}$ and $D_{Chi}$ contain 49.0 and 55.8 million tokens, respectively (obtained by the tokenizer of CodeLlama). The statistics of the dataset can be accessed in Table \ref{tab:statistic}. For the reasons for choosing Chinese as the studied natural language, readers can refer to Section \ref{sec:threat}. 

\begin{table*}[ht]
    \centering
    \caption{Statistics and programming language distribution of our constructed dataset \data.}
    \begin{tabular}{c|c|cccccccccc|c}
    \toprule
    Version & \#Tokens & Python & JavaScript & SQL & Java& C++ & HTML & CSS & PhP & Bash & Others & Total \\
    \midrule
    $D_{Eng}$ & 49.0M& \multirow{2}{*}{49,346} & \multirow{2}{*}{10,291} & \multirow{2}{*}{8,105} & \multirow{2}{*}{6,976} & \multirow{2}{*}{3,597} & \multirow{2}{*}{1,879} & \multirow{2}{*}{1,800} & \multirow{2}{*}{1,286} & \multirow{2}{*}{814} & \multirow{2}{*}{7,672} & \multirow{2}{*}{91,766} \\
    $D_{Chi}$ & 55.8M & & & & & & & & & & & \\
    \bottomrule
    \end{tabular}
    
    \label{tab:statistic}
\end{table*}

\subsubsection{Exploration on Mitigating Multi-NL Understanding Bias}
To study the impact of used data and training methods of instruction tuning on the multi-NL bias of LCMs, we explore four instruction tuning methods
based on the constructed instruction tuning dataset. 

\begin{enumerate}
    \item \textbf{English-Based Tuning.} We use the English version of our \data $D_{Eng}$ and construct the training instance as the format ``$\mathit{[INS]\backslash n[ANS]}$'' to train LCMs, where \textit{[INS]} and \textit{[ANS]} represent the instruction and corresponding answer, respectively.
    \item \textbf{Chinese-Based Tuning.} We use the translated dataset $D_{Chi}$ to train LCMs in the same way as English-based tuning.
    \item \textbf{Mixed-NL-Based Tuning.} We randomly sample 50\% training instances from $D_{Eng}$ and $D_{Chi}$ with the same data construction.
    \item \textbf{Translation-Aware Tuning.} We propose to take both the English and Chinese instructions into account. Specifically, we construct the training data through a translation-aware template that provides both English instruction, translated Chinese instruction, and the corresponding Chinese answer. The details of the template for translation-aware tuning are shown in our anonymous repository.
\end{enumerate}


\subsubsection{Exploration on Mitigating Multi-PL Generation Bias}

We further investigate the impact of programming languages used for instruction tuning on the code generation performance and multi-PL bias of LCMs.

Specifically, we split the dataset into two parts according to different PLs including Python and Other-PLs due to the imbalanced PL distribution in our \data. Then we conduct instruction tuning in three methods including 1) \textbf{Python-based Tuning}, 2) \textbf{Other-PLs-based Tuning}, and 3) \textbf{Full data-based Tuning}, respectively. In methods 1) and 2), we utilize the Python and Other-PLs parts to tune LCMs, respectively. In the method 3), we conduct instruction tuning with all training instances in \data. 

\section{Experimental Setup}\label{sec:setup}

\subsection{Selected LCMs}
In this paper, we select three kinds of popular and state-of-the-art LCMs
with their versions in different sizes. In specific, our selected LCMs are:
\begin{itemize}
    \item \textbf{StarCoder} \cite{li2023starcoder} is a large language model trained on the mixture of source code and natural language texts. Its training data incorporate more than 80 different programming languages as well as text extracted from GitHub issues and commits and from notebooks. The total account of training tokens exceeds 1T. We select its 3B, 7B, and 16B versions in our experiments.
    \item \textbf{CodeLlama} \cite{codellama} is a family of large language models for code based on LLama 2~\cite{Touvron2023Llama2O} with state-of-the-art code generation, blank infilling, and long-context processing capabilities. In this paper, we choose CodeLlama's base model (i.e., CodeLlama Base) in three different sizes including 7B, 13B, and 34B for instruction tuning. 
    \item \textbf{DeepSeek-Coder} \cite{guo2024deepseek} is a series of large code models that have an identical architecture to CodeLlama. DeepSeek-Coder is trained from 2T tokens from scratch, which comprises 87\% code and 13\% natural language in both English and Chinese. DeepSeek-Coder achieves state-of-the-art performance in a variety of code intelligence tasks. Specifically, we choose DeepSeek-Coder Base in sizes of 1.3B, 6.7B, and 33B in this paper.
\end{itemize}

\subsection{Evaluation Metrics}
Following the prior studies \cite{shen2023pangu, li2023starcoder, codellama}, we use the Pass@k metric to evaluate the accuracy of LCMs in solving programming problems, examining whether LCMs can pass all unit tests within $k$ solutions. The metric can be described as the following
\begin{equation}
 \mathit{pass@k}= \frac{\sum \limits_{i=1}^{n}\prod_{j=1}^k(\mathbbm{1}(pass_{s_i^j})}{n}
\end{equation}
where $n$ and $k$ denote the number of problems and the number of generated solutions, respectively. $s_i^j$ indicates the $j$-th solution for the $i$-th problem. The function $\mathbbm{1}(x)$ returns 1 if $x$ is True and otherwise returns 0, and $pass(s)$ returns True if the solution $s$ can pass all unit tests. In this paper, following previous work \cite{wei2023magicoder, li2023starcoder, luo2023wizardcoder}, we choose Pass@1 as our metric, i.e., $k=1$.
\begin{table}[t]
    \centering
    \caption{Hyper-parameter settings.}
    \resizebox{\linewidth}{!}{
    {\begin{tabular}{c|c||c|c}
    \toprule
       Hyperparameter  & Value &  Hyperparameter  & Value\\
     \midrule
      Optimizer & AdamW\cite{loshchilov2018decoupled} & Warm-up steps   & 100 \\
      Learning rate & 5e-6 & Training batch size & 512 \\
      LR scheduler& Cosine Scheduler \cite{loshchilov2016sgdr}& Validation batch size & 32 \\
        Sequence Len. & 2,048 &  Adam epsilon & 1e-8 \\
      Max. gradient norm & 1.0 & Precision & BF16\\
      \midrule
      Max Gen. Tokens & 512 & Top-P & 0.95 \\
      \bottomrule
    \end{tabular}
    }
    }
    \vspace{-6pt}
    \label{tab:param}
\end{table}

\subsection{Implementation Details}

All the experiments are run on a server with 8*A100 GPU with 80GB graphic memory. Specifically, we utilize Optimizer State Sharding (ZeRO 3) techniques in DeepSpeed \cite{ren2021zero, rajbhandari2020zero} to save GPU memory and improve training efficacy. For larger models such as CodeLlama 34B and DeepSeek-Coder 33B, we additionally offload the optimizer state into CPU memory to further avoid out-of-CUDA memory issues. The training hyper-parameters are listed in Table \ref{tab:param} following previous work \cite{li2023starcoder,luo2023wizardcoder, wei2023magicoder}. We randomly select 5\% of the training samples as a validation set and use the checkpoint with the lowest validation loss for inference. 

For fast inference, we utilize vLLM \cite{kwon2023efficient} based on PagedAttention to improve efficiency. The inference hyper-parameter is also listed in Table \ref{tab:param}. After generation, we conduct post-processing (e.g., truncating the content beyond the solution function) to ensure the generated code can be correctly evaluated following previous work \cite{li2023starcoder,allal2023santacoder}.

\section{Experiment Analysis}
\label{sec:results}
In this section, we elaborate on the answers to the proposed research questions based on the experimental results. 

\begin{table*}[t]
    \centering
    \caption{Code generation results (Pass@1) of the multi-NL and multi-PL benchmarks on \textit{X-HumanEval-X}, where SC, CL, and DSC indicate StarCoder, CodeLlama, and DeepSeek-Coder, respectively. The column \textbf{NL-Bias} and \textbf{PL-Bias} denote the maximum performance gap of the average Pass@1 among different NLs and PLs. ``-" denotes that the instruction tuning version of the models is not released.}
    \resizebox{\linewidth}{!}{
    \begin{tabular}{c|c|ccccccccc|ccc}
    \toprule
      Benchmark-PL & Benchmark-NL &SC-3B & SC-7B& SC-15B & CL-7B & CL-13B & CL-34B& DSC-1.3B & DSC-6.7B & DSC-33B & Avg& NL-Bias & PL-Bias\\
      \midrule
      \multicolumn{14}{c}{Base Model} \\
      \midrule
     \multirow{2}{*}{Python}  & English & 22.56 & 26.82 & 31.70 & 31.09 & 35.36 &54.87& 28.65 & 49.39 & 55.48 & 37.32 & \multirow{2}{*}{$\Delta 17.25\%$ }& \multirow{6}{*}{$\Delta 23.70\%$}\\
     & Chinese & 18.29 & 23.78& 26.21 & 29.26 & 28.04 & 41.46& 25.61 & 46.34 & 47.56 & 31.84& &\\
     \cline{1-13}
     \multirow{2}{*}{C++} & English & 19.51 & 24.39& 29.87 & 28.04 & 34.75 & 50.60& 31.70 & 43.90 & 51.21 &34.88&\multirow{2}{*}{$\Delta 13.65\%$} & \\
     & Chinese & 18.29 & 22.56& 26.21 & 21.95 & 30.49 & 40.24 & 31.09 & 39.02 & 46.34& 30.69& & \\
     \cline{1-13}
     \multirow{2}{*}{Java} & English&20.73& 22.56& 24.39 & 25.00 & 30.49 & 50.00 & 30.49 & 46.95 &  50.60& 33.47 & \multirow{2}{*}{$\Delta 30.03\% $} & \\
     & Chinese &15.85&20.12 & 19.51 & 18.90 & 23.78 & 31.09 & 22.56 & 36.58 & 43.29& 25.74 & &\\
     \midrule
     \multicolumn{14}{c}{Instruction Tuned Model} \\
     \midrule
     \multirow{2}{*}{Python}  & English& - &- & 34.15 & 35.97 & 42.68 & 53.04& 59.14 & 71.95 & 73.17 & 52.87&\multirow{2}{*}{$\Delta 14.31\% $} & \multirow{6}{*}{$\Delta 13.04\%$}\\
     & Chinese& - &- & 29.26 & 31.09 & 37.19 &45.12  & 54.87 & 62.19 & 64.02 &46.25 & &\\
     \cline{1-13}
     \multirow{2}{*}{C++} & English& - &- & 28.66 & 32.92 & 42.07 & 47.56& 48.17 & 62.80 & 65.24 & 46.77 &\multirow{2}{*}{$\Delta13.05\%$ } & \\
     & Chinese& - &- & 25.61 & 30.49 & 30.49 & 44.51 & 45.12 & 55.48 & 57.92 & 41.37 &  &\\
     \cline{1-13}
     \multirow{2}{*}{Java} & English& - &- & 29.26 & 39.63 & 37.19 & 52.49 & 56.70 & 71.34 & 73.17& 51.40& \multirow{2}{*}{$\Delta16.39\%$} & \\
     & Chinese& - &- & 25.00 & 31.09 & 28.04 & 39.63 & 50.60 & 65.85 & 68.90&44.16 & & \\
     \bottomrule
     
    \end{tabular}
    }
    \label{tab:pre}
    \vspace{-6pt}
\end{table*}

\subsection{RQ1: Existence of Multi-Lingual Bias in LCMs}

Given a programming problem, we expect LCMs to generate correct solutions no matter the language that instructions are described in such as English and Chinese. In addition, we also expect LCMs to be able to solve programming problems in different PLs. Therefore, in RQ1, we evaluate the code generation capabilities of LCMs and the potential multi-lingual bias across different PLs including Python, C++, and Java in different NLs, including English and Chinese.
To answer RQ1, we utilize our constructed \textit{X-HumanEval-X} to evaluate the performance of the selected nine LCMs and compare their performance across different NLs and PLs. 
Besides base models (i.e., versions after pre-training), their corresponding instruction-tuned models (i.e., versions after instruction tuning) are also evaluated.
The results are shown in Table \ref{tab:pre}. From the results, we can achieve
the following observations. 

1) \textbf{LCMs
exhibit bias in multi-NL understanding.} All the experimented LCMs exhibit a notable multi-NL bias, i.e., the performance gap of LCMs when generating code with instructions in English and Chinese, across all programming languages. More precisely, when instructions are presented in Chinese, the average Pass@1 rate for the LCMs under study drops by 17.2\% and 14.3\% for the base and instruction-tuned model versions in Python, respectively.  In the case of the base model version of CodeLlama 34B, its capability to generate code in Java experiences an even more pronounced decrease, with a performance reduction of 37.8\%.  The results reveal a pronounced bias in LCMs regarding their ability to understand different natural languages when tasked with code generation.


2) \textbf{LCMs have biased abilities in the multi-PL generation. } From the multi-PL perspective, we observe that LCMs perform variously in generating solutions in different programming languages. For instance, the base model versions of LCMs achieve the best Pass@1 rate in the Python language, which is 5.7\% and 11.3\% higher than that in C++ and Java, respectively. The code generation performance in the Java language is the lowest among the three experimented programming languages with both English and Chinese instructions. Compared to Python, the average Pass@1 rates for Java programming problems, when instructions are provided in English and Chinese, are lower by 11.5\% and 23.7\%, respectively.  We attribute the bias of different PLs on base models to the training objective of base models, which is focused on auto-aggressive decoding. Without instruction tuning, base models present unsatisfactory capabilities to follow instructions and lead to a scenario where base models struggle with determining ``when to stop" during code generation.  This problem is specifically severe when generating solutions in Java, primarily because solutions in Java are typically organized as classes rather than functions. Thus, base models achieve the worst performance in Java on average. 

However, the models after instruction tuning (the lower part of the table) exhibit their poorest performance in C++ (average 46.77\% in English instructions), which presents the maximum performance gap at 13.04\%. This discrepancy can be ascribed to the enhanced capability of LCMs to learn to follow instructions and generate solutions with the proper format, obtaining larger improvements in Java (achieving the average Pass@1 at 51.4\% in English). We suppose that
the inherent complexities associated with C++ programming result in LCMs exhibiting inferior performance when compared to their counterparts in Java and Python.

\finding{1}{Current LCMs
exhibit pronounced bias in both multi-natural language understanding and multi-programming language generation. Specifically, when transitioning from English to Chinese instructions, the average Pass@1 rate experiences a minimum decrease of 13\%. When generating source code across different programming languages, multi-PL bias reaches as high as 23.7\%. }

\subsection{RQ2: Mitigating Multi-Lingual Bias in Prompting}
\begin{table*}[t]
    \centering
    \caption{Code generation results (Pass@1) of different prompting strategies for mitigating multi-NL bias.}
    \begin{tabular}{c|ccccccccc|c}
    \toprule
      Methods &SC-3B & SC-7B& SC-15B & CL-7B & CL-13B & CL-34B& DSC-1.3B & DSC-6.7B & DSC-33B & Avg\\
      \midrule
      Chinese (w/o Trans.) & 18.29 & 23.78& 26.21 & 29.26 & 28.04 & 41.46& 25.61 & 46.34 & 47.56 & 31.84 \\
      \midrule
      Self Trans. & 1.83 & 2.44 & 8.54 & 7.32 & 11.59 & 29.26& 7.32 & 13.41 & 26.21& 11.99 \\
      One-Step Trans. & 20.12&  25.61& 26.82 & 28.04 & 32.31 &48.17 & 28.62 & 45.73 & 48.17 & 33.73 \\
      Multi-Step Trans. & 21.95 & 25.61& 28.66 & 29.26 & 32.92 & 52.44 & 30.49 & 47.56 & 54.87 & 35.97 \\
      \midrule
      Original English& 22.56 & 26.82 & 31.70 & 31.09 & 35.36 &54.87& 28.65 & 49.39 & 55.48 & 37.32 \\
      \bottomrule

    \end{tabular}
    
    \label{tab:pre2}
\end{table*}

To answer the second RQ, we investigate the effectiveness of three prompting strategies including self-translation, one-step translation, and multi-step translation in mitigating multi-NL bias. The experiment results are shown in Table \ref{tab:pre2}.

From the table, we can find that in the case of self-translation, a notable decline in the performance is observed. Specifically, the average Pass@1 of code generation through self-translation stands at merely 12.0\%, marking a pronounced decrease of 62.3\% compared to performance with original Chinese instructions. These unfavorable outcomes indicate that current LCMs, primarily trained on source code, exhibit inadequate translation capabilities, making self-translation even enlarge the multi-NL bias.

For one-step and multi-step translation, these strategies achieve a substantial improvement in mitigating the multi-NL bias. Specifically, the average Pass@1 rate increases by 5.9\% and 13.0\% compared to the original Chinese instructions, reducing the multi-NL bias from the original 17.2\% to 10.6\% and 3.8\%, respectively. Such results demonstrate that effectively translating Chinese instructions into English helps mitigate the bias in multi-NL understanding. 
Furthermore, our observations indicate that multi-step translation substantially outperforms one-step translation, effectively narrowing the performance gap and approximating
the performance in the original English instructions. This disparity can be attributed to the limitations of current translation tools in handling content that intertwines textual and symbolic elements,
leading to a compromise in both the fidelity of translation and the efficacy of mitigating multi-NL bias.

\finding{2}{Prompting LCMs through one-step and multi-step translation can mitigate the multi-NL bias, reducing the bias from 17.2\% to as low as 3.8\%. However, for self-translation, due to the unsatisfactory translation capabilities of LCMs, the average Pass@1 drops by 62.3\%.}

\subsection{RQ3: Mitigating Multi-Lingual Bias in Instruction Tuning}
In RQ3, we opt to focus on the base model versions of the LCMs because the specifics of their instruction tuning phase, including the data and training strategies employed, are not accessible to us. Using base models allows us to concentrate solely on understanding how the choice of data and training methods used for instruction tuning influences the model performance.
\subsubsection{RQ3.1: Multi-NL
Understanding Bias}

We conduct instruction tuning on the selected LCMs through the proposed four training methods in Section \ref{sec:moti} and the results are shown in Table \ref{tab:rq1}. 

For \textbf{English-based Tuning}, we can observe that instruction tuning with pure English training instances can notably improve the code generation performance with both English and Chinese instructions. Specifically, compared to the original base models, instruction tuning by English data achieves an average of 28.2\% and 34.5\% improvement on the Pass@1 metric with
English and Chinese instructions, respectively. The results reveal inter-dependencies within the knowledge representations of various natural languages embedded in LCMs. The inter-dependencies render that
instruction tuning LCMs in one natural language can concurrently enhance their performance across different languages.

However, despite the noticeable performance improvement, an obvious disparity persists between the models' treatment of English and Chinese instructions.
After English-based tuning, the average Pass@1 rate of Chinese instructions is 11.75\% lower than that of English, implying that relying solely on English data cannot well
address the underlying
bias towards non-English languages. 
\begin{table*}[t]
    \centering
    \caption{Comparison of code generation results (Pass@1) of English and Chinese benchmarks on the \textit{X-HumanEval-X}. SC, CL, and DSC indicate StarCoder, CodeLlama, and DeepSeek-Coder, respectively. All experiments are conducted in Python. The column \textbf{NL-Bias} denotes the maximum performance gap of the average Pass@1 among different NLs. \textbf{Bold} and \underline{underline} results represent the best performance in English and Chinese benchmarks, respectively.}
    \resizebox{\linewidth}{!}{
    \begin{tabular}{c|c|ccccccccc|cc}
    \toprule
      Training Method & Benchmark-NL &SC-3B & SC-7B& SC-15B & CL-7B & CL-13B &CL-34B & DSC-1.3B & DSC-6.7B & DSC-33B &Avg & NL-Bias\\
      \midrule
      \multirow{2}{*}{Base Models} & English & 22.56 & 26.82	&31.70 &31.09	&35.36	& 54.87 &	28.65 & 49.39 & 55.48 & 37.32 & \multirow{2}{*}{$\Delta 17.25\%$}\\
      & Chinese & 18.29 & 23.78 & 26.21& 29.26& 28.04& 41.46& 25.61 & 46.34 & 47.56 & 31.83 & \\
      \midrule
      \multirow{2}{*}{English-based} & English & 30.49&34.14& 44.51& 44.51 &	52.49&	62.80 &38.41	&\textbf{59.14}& 64.02& 47.83 & \multirow{2}{*}{$\Delta 11.75\%$}\\
      & Chinese & 28.65 & 30.49 & 35.36& 39.02 & 46.34 & 56.70& 34.14& 54.78 & 59.75 & 42.80 & \\
      \midrule
      \multirow{2}{*}{Chinese-based} & English & 30.49 &36.58 & \textbf{45.12}& 43.29	&54.26	& 62.80&34.14& 55.48 &62.80 & 47.22 & \multirow{2}{*}{$\Delta 2.79\%$}\\
      & Chinese& \underline{32.31}& 33.63& 38.41& 41.46& 49.35&59.14 & \underline{37.19}& \underline{57.92} &64.02 & 45.94 & \\
      \midrule
      \multirow{2}{*}{Mixed-NL-based} & English & 32.31 & 36.58& 43.29& \textbf{46.34}& 54.26& 63.41& 35.36& \textbf{59.14} & 64.02& 48.30 & \multirow{2}{*}{$\Delta 6.11\%$}\\
      & Chinese & 31.70 & \underline{34.14}& \underline{39.63}& 41.46& 47.56& 58.53& 35.97& 57.31 & 63.41& 45.52 & \\
      \midrule
      \multirow{2}{*}{Translation-aware} & English & \textbf{34.13}& \textbf{38.41}& \textbf{45.12}& 45.12& \textbf{55.48}	& \textbf{64.02}&\textbf{39.02} & 58.53 & \textbf{65.24} & \textbf{49.45} & \multirow{2}{*}{$\Delta 6.09\%$}\\
      & Chinese& \underline{32.31}& \underline{34.14}& \underline{39.63}& \underline{42.07}& \underline{51.21}& \underline{60.97}& 36.58& 57.31& \underline{65.24} & \underline{46.61} & \\
      \bottomrule

    \end{tabular}
    }
    \label{tab:rq1}
\end{table*}

\finding{3}{Instruction tuning LCMs with English dataset can bring substantial performance improvement on code generation in both English and Chinese instructions, i.e., the Pass@1 rate improves 28.2\% and 34.5\%, respectively. However, the bias between English and Chinese benchmarks is still severe, indicated by the obvious performance gap at 11.75\%.}

For \textbf{Chinese-based Tuning}, similar to purely English tuning, we also observe a considerable improvement in the performance of the
LCMs, i.e., the Pass@1 rates obtain an average relative improvement of 26.6\% and 44.3\% in English and Chinese instructions, respectively. This observation aligns well with the English-based tuning. The difference is that by utilizing Chinese data, LCMs register a 7.3\% larger improvement on the Chinese benchmark while suffering a decrease of 1.3\% on the English benchmark compared to English-based tuning.  For the overall performance of the two languages, Chinese-based tuning outperforms English-based tuning by 2.8\% in terms of the average Pass@1 rates of the two benchmarks. The results indicate that compared with English-based tuning, Chinese-based tuning can achieve competitive performance on the English benchmark while performing substantially better on the Chinese benchmark.

From the perspective of the performance gap between English and Chinese, even though the average performance in English is still better than that in Chinese, we observe that tuning with Chinese leads to a gap of 2.79\%, which is 76\% lower than that of English-based tuning. Specifically, for DeepSeek-Coder after Chinese-based tuning, the average Pass@1 rates of the Chinese benchmark even outperform that of the English benchmark (e.g., 37.19 v.s. 34.14 in DeepSeek-Coder 1.3B). We attribute this superior performance under Chinese instructions to the deliberate inclusion of a non-trivial amount of Chinese corpora during the pre-training phase of DeepSeek-Coder \cite{guo2024deepseek} for improving the models' comprehension of Chinese. 

\begin{table*}[t]
    \centering
    \caption{Comparison of code generation results (Pass@1) in Python, C++, and Java benchmarks. SC, CL, and DSC indicate StarCoder, CodeLlama, and DeepSeek-Coder, respectively. All experiments are conducted in English. The column \textbf{PL-Bias} denotes the maximum performance gap of the average Pass@1 among different PLs. \textbf{Bold}, \textit{italic} and \underline{underline} results represent the best performance in Python, C++, and Java benchmarks, respectively.}
    \resizebox{\linewidth}{!}{
    \begin{tabular}{c|c|ccccccccc|cc}
    \toprule
      Training Method & Benchmark-PL &SC-3B & SC-7B& SC-15B & CL-7B & CL-13B &CL-34B & DSC-1.3B & DSC-6.7B & DSC-33B &Avg & PL-Bias\\
      \midrule
      \multirow{3}{*}{Base Models} & Python & 22.56 & 26.82 & 31.70 & 31.09 & 35.36 & 54.87 & 28.65 & 49.39 & 55.48 & 37.32 & \multirow{3}{*}{$\Delta 11.50\%$}\\
      & C++ & 19.51 & 24.39& 29.87 & 28.04 & 34.75 & 50.60& 31.70 & 43.90 & 51.21 &34.88 &\\
      &Java &  20.73& 22.56& 24.39 & 25.00 & 30.49 & 50.00 & 30.49 & 46.95 &  50.60& 33.47 & \\
      \midrule
      \multirow{3}{*}{Python} & Python & \textbf{30.49} & 33.53 & 43.29 & \textbf{44.51} & 51.21 &  \textbf{62.80}& \textbf{38.41} & \textbf{59.75} & 63.41 & 47.49 &\multirow{3}{*}{$\Delta 10.05\%$}\\
      & C++ & 21.95 & 31.70 & 39.02 & 37.80 & 46.95 & 57.92 & \textit{37.80} & \textit{57.31} & 57.92 & 43.15 &\\
      & Java & 28.04 & 31.70 & 39.63 & 39.63 & 52.43 & 57.92 & 37.80 & 54.26 & 58.53 & 44.44 &\\
      \midrule
      \multirow{3}{*}{Other-PLs} & Python &29.26 & 32.92 & 37.80 & 34.75 & 47.56 & 60.97 & 37.80 & 57.92 & 61.58 & 44.51 &\multirow{3}{*}{$\Delta 6.73\%$} \\
      & C++ & 23.78 & 33.53 & 39.02 & 37.80 & 50.00 & 56.70 & 35.36 & 54.87 & 57.92 & 43.22 &\\
      & Java & 29.87 & 33.53 & 41.46 & 40.24 & 49.35 & 58.53 & 40.85 & \underline{57.92} & \underline{63.41} & 46.13 &\\
      \midrule
      \multirow{3}{*}{Full data} & Python & \textbf{30.49} & \textbf{34.14} & \textbf{44.51} & \textbf{44.51} & \textbf{52.49} & \textbf{62.80} & \textbf{38.41} & 59.14 & \textbf{64.02} & \textbf{47.83} &\multirow{3}{*}{$\Delta 5.97\%$}\\
      & C++ & \textit{26.21} & \textit{34.14} & \textit{41.46} & \textit{41.46} & \textit{51.52} & \textit{59.14} & \textit{37.80} & 56.70 & \textit{59.14} & \textit{45.29} &\\
      & Java & \underline{32.31} & \underline{35.36} & \underline{43.29} & \underline{44.51} & \underline{55.48} & \underline{60.97} & \underline{41.46} & 57.31 & 62.80 & \underline{48.17} &\\
     \bottomrule
     
    \end{tabular}
    }
    \label{tab:rq2}
\end{table*}
\finding{4}{Compared to English-based tuning, Chinese-based tuning improves the Pass@1 rate by 7.3\% in Chinese instructions while sacrificing 1.3\% in English instructions, achieving a lower multi-NL bias at 2.79\%. }

For \textbf{Mixed-NL-based Tuning}, we construct a mixed-NL dataset by randomly selecting 50\% training instances from $D_{eng}$ and $D_{chi}$, respectively. 
From the table, we observe that in the English benchmark, mixed-NL-based tuning performs better than both English and Chinese-based tuning, obtaining an average of 1.0\% and 2.2\% improvement on overall Pass@1, respectively. Such improvements indicate that the capabilities of LCMs in English can be further boosted by involving Chinese data. For the Chinese benchmark, the average Pass@1 of mixed-NL-based tuning is slightly lower than that of purely Chinese tuning (0.9\%), which still outperforms English-based tuning by 6.4\%. The results demonstrate that mixing English and Chinese can further boost LCMs in English instructions and obtain promising capability in Chinese instruction understanding, leading to a multi-NL bias at 6.11\%. This dual-language training method yields superior overall performance compared to training exclusively in a single language.


Through \textbf{Translation-Aware Tuning}, the average multi-NL bias achieves 6.09\%, which is close to that of mixed-NL-based tuning. In addition, LCMs achieve the best performance in both English and Chinese benchmarks. Specifically, translation-aware tuning improves the average Pass@1 by 2.38\% and 2.95\% over mixed-NL tuning in the English and Chinese benchmarks, respectively.
The notable performance improvement can be ascribed to the LCMs' ability to discern and learn the intricate triplet relationships among English instructions, Chinese instructions, and their corresponding responses during the instruction tuning phase. 

\finding{5}{For mixed-NL-based tuning and translation-aware tuning,
LCMs present similar multi-NL bias at 6.1\%. From the perspective of code generation capabilities, LCMs achieve better performance over tuning with exclusively English or Chinese data. }

\subsubsection{RQ3.2: Multi-PL Generation Bias}
In this research question, we conduct a quantitative analysis to explore the influence of instruction tuning with various programming languages on the performance of LCMs. We split the training dataset based on the associated programming languages and their training instances, specifically categorizing it into Python and Other-Programming Languages (Other-PLs) following previous work \cite{wei2023magicoder}, comprising 49,346 and 42,420 instances respectively. Subsequently, these segmented datasets are employed to train LCMs. Then we assess the LCMs' code generation performance across various programming languages, including Python, Java, and C++. We present our results in Table \ref{tab:rq2}.

For \textbf{Python-based Tuning}, from the table, we can observe that training on the data of a single Python language brings remarkable performance improvement among the benchmarks. Specifically, the average Pass@1 of LCMs increases by 27.3\%, 23.7\%, and 32.8\% in Python, C++, and Java, respectively. This observation about programming languages aligns with the natural languages, indicating an implicit relationship among the knowledge about programming languages embedded in LCMs. Through instruction tuning LCMs with even a single language, the capabilities in other languages will also be exploited (no matter a natural language or programming language). This improvement suggests that
the instruction tuning process, even when it is concentrated on a specific language, can enhance the multi-lingual generalization, boosting the models' overall ability to understand instructions and generate code.

From the perspective of multi-PL bias, despite the substantial improvement, multi-PL bias is still obvious, i.e., a 10.05\% performance gap between Python and C++ benchmarks is observed.



After \textbf{Other-PLs-based Tuning}, LCMs enhance their code generation capabilities across multiple programming languages, although the degree of improvement varies when compared to training exclusively on the Python language dataset. Specifically, when the training is conducted on programming languages other than Python (Other-PLs), there is an observed increase in the Pass@1 metric by 19.3\%, 23.9\%, and 37.8\% for the three benchmarks, respectively. Compared to Python-based tuning, training on other PLs improves the performance in Java by 3.8\% while suffering a 6.5\% reduction for Python. In addition, the method utilizes more PLs for instruction tuning, mitigating the multi-PL bias compared to original base models and Python-based tuning by at least 30\% (i.e., the performance gap between C++ and Java is 6.7\%).
These outcomes suggest that the selection of PLs can bring greater improvement in LCMs' capabilities to generate the corresponding PL, thus, the limited diversity of PLs used for instruction tuning tends to be insufficient to mitigate multi-PL bias.

\finding{6}{The code generation performance in multi-programming languages can be improved by instruction tuning, regardless of the specific programming language used for training. Notably, more substantial improvements are observed when LCMs are tuned with datasets corresponding to the target programming language, resulting in a multi-PL bias larger than 6.7\%.}

For \textbf{Full Data-based Tuning} with the mixture of the above-mentioned PLs, LCMs achieve the best average Pass@1 in all PLs including Python, C++, and Java among experimented methods. Compared to the base models, full data-based tuning increases the overall Pass@1 by 27.5\%, 25.7\%, and 39.1\% in the three experimented benchmarks, respectively.  Such superior performance of full data-based training demonstrates that increasing the diversity from the aspect of programming languages can further bring enhancement in the code generation capabilities of multiple programming languages.

In terms of the bias among experimented PLs, 
we observe that increasing the diversity of PLs for instruction tuning aids in mitigating bias in multi-PL generation.  For instance, when training is conducted exclusively with Python, the widest performance gap observed is 10.05\% (between Python and C++). However, when tuning is performed using a full dataset that encompasses multiple programming languages, this gap narrows by 40\%  (i.e., a bias of 5.97\% between Java and C++).

\finding{7}{Full data-based tuning that further enhances the diversity of PLs mitigates the average multi-PL bias by 40\% compared to Python-based tuning. In addition, the average Pass@1 rate obtains the largest improvement at 27.5\%, 25.7\%, and 39.1\% in Python, C++, and Java, respectively.
}

\section{Discussion}\label{sec:discuss}

\subsection{Implication of Findings}
In this section, we discuss the implications of our work for researchers and developers.

\textbf{For researchers.} Our research demonstrates that current large code models present obvious multi-lingual bias in the code generation task. With well-designed training methods and dataset construction, instruction tuning substantially improves LCMs' performance of code generation and mitigates the multi-lingual bias. However, as shown in the results of RQ3, the multi-lingual bias still persists in current LCMs. Our results also reveal the potential research directions in the era of LCM for the community. Specifically:

\begin{itemize}
    \item \textbf{Exploring to collect high-quality multi-lingual data.} 
    The users in current software communities such as StackOverflow mainly concentrate on certain NLs (i.e., English) or PLs (e.g., Python and JavaScript) \cite{userstatis}, which potentially leads to a degraded diversity when collecting data from these communities.
    Therefore, exploring effective methods for gathering, evaluating, and integrating diverse language data for training LCMs is crucial. 
    \item \textbf{Exploring to amplify underrepresented languages.} In addition, the highly imbalanced multi-lingual data used for training, e.g., the imbalanced distribution of programming languages as reported in \cite{lozhkov2024starcoder}, result in the under-representation of some languages. Thus, developing strategies that can effectively amplify the presence and influence of underrepresented languages in the training data (e.g., weighted training \cite{wang2022label} and language-agnostic representations \cite{zugner2020language}) needs to be further investigated. 
    \item \textbf{Exploring more sophisticated training methods.} This paper aims to mitigate the multi-lingual bias from the angle of instruction tuning. Besides instruction tuning, the exploration of more sophisticated fine-tuning methodologies (e.g., multi-lingual preference learning \cite{christiano2017deep, rafailov2024direct}) presents a promising avenue for future research.

    
\end{itemize}

\textbf{For developers.} Instruction tuning enables the pre-trained LCMs to follow human instructions and better exploit the knowledge embedded in the model. Our findings indicate that the multi-lingual data and training methods have a substantial impact on the performance of LCMs. Based on our findings, we conclude the following insights and takeaways for developers to conduct instruction tuning and adapt LCMs into practice in their work.

\begin{itemize}
    \item The presence of multi-lingual bias in current LCMs can affect their performance in generating source code from instructions given in languages other than English.  This bias may lead developers, facing unsatisfactory outcomes with their native or preferred languages, to resort to using English when interacting with LCMs.
    \item Instruction tuning LCMs with a single language (no matter NL and PL) improves the performance of LCMs in code generation across other languages.
    \item Instruction tuning helps to mitigate the multi-lingual bias. However, tuning with a purely English dataset still results in a notable degree of bias, specifically measured at 11.75\%.
    \item Enhancing the diversity from both NL and PL perspectives of the instruction tuning dataset further improves the generation capabilities of LCMs and mitigates multi-lingual bias, making LCMs more accessible and useful to global developers.

\end{itemize}

\subsection{Prompting or Instruction Tuning}
In our experiments in RQ2 and RQ3, we demonstrate that the multi-NL bias can be potentially mitigated in both the prompting and instruction tuning phases. In this section, we discuss the advantages and limitations of the two methods. 

For \textbf{prompting}, we observe that one-step translation and multi-step translation aid in mitigating the multi-NL bias by 40\% and 76\%, respectively. Despite the contribution to mitigating multi-NL bias, these methods have the following limitations: 1) The benefits come with auxiliary translation tools and non-trivial costs. For example, in the multi-step translation process, we averagely translate 3.9 statements for each programming problem. 
2) Translation helps to mitigate the bias by narrowing the performance gap between Chinese and English instructions; however, the essential code generation capabilities of LCMs are not improved. 3) LCMs typically generate responses such as code comments and explanations in the same language as the input. Consequently, when users who are not proficient in English opt to translate their queries into English, they may find the English responses difficult to comprehend, resulting in additional costs to translate back.

For \textbf{instruction tuning}, as noted in RQ3, we observe that instruction tuning not only effectively mitigates the multi-NL bias by as high as 84\% but also substantially improves the code generation capabilities of LCMs, as evidenced by an overall 39\% increase in the Pass@1 rate. However,  instruction tuning involves the creation of a multi-lingual dataset and demands computational resources for training the LCMs. This requirement could pose practical challenges and resource constraints in implementation.

Drawing upon our empirical findings, developers can choose proper methods to mitigate the multi-lingual bias according to their resources and expectations in practice.
\subsection{Threats to Validity}\label{sec:threat}
We have identified the following major threats to validity:

\textbf{Limited LCMs}. The experiments are based on open-source popular LCMs,
which may bring bias in the results. The reason we refrain from utilizing close-source models such as ChatGPT \cite{chatgpt} and Gemini \cite{gemini} stems from the fact despite their superior capabilities, closed-source models are prohibited within certain companies due to privacy concerns \cite{wang2023practitioners}. Consequently, instruction tuning open-source LCMs as an internal programming assistant presents a viable solution, which is the point that this paper focuses on. To mitigate this issue, we select three types of popular LCMs of varying sizes, ranging from 1.3 billion to 34 billion parameters to control the threat.

\textbf{Limited natural languages}. In this paper, we specifically focus on English and Chinese for conducting experiments on multi-NL understanding. This choice is grounded in the fact that English and Chinese are recognized as the two most prevalent languages globally \cite{nlstatis}. Additionally, according to existing studies \cite{Wang2023AllLM}, Chinese and English present a high imbalance from the perspective of training data. In addition, despite the prevalence of Chinese, a substantial performance disparity with English is still observed. Thus, we believe the findings obtained by investigating the multi-NL bias in Chinese and English can also be generalized to other languages. 
\section{Related Work}\label{sec:related}

\subsection{Large Code Models}

The use of large language models in natural language processing inspires researchers and companies to develop LCMs for programming tasks. Generally, LCMs can be obtained through two approaches: continuing the training of foundation LLMs or pre-training from scratch \cite{li2023starcoder, allal2023santacoder, codex, zheng2023codegeex, codellama, guo2024deepseek}. CodeLlama~\cite{codellama} is an example of the former, which is based on the LLaMA2 foundation model~\cite{touvron2023llama2}. On the other hand, Starcoder2~\cite{lozhkov2024starcoder} is an example of the latter, trained from scratch with over 600 programming languages. 
In addition to these open-source models, big companies also develop their coding products with LCMs, such as GitHub Copilot~\cite{copilot} and Tabnine~\cite{tabnine}. LCMs show promising results in various code tasks, including code completion and summarization~\cite{nijkamp2022codegen}. However, they face challenges similar to LLMs, such as safety output~\cite{li2022cctest} and intellectual property concerns~\cite{li2023protecting,li2023feasibility}.

Apart from source-level code, LCMs have also been designed for low-level code, such as decompilation and LLVM IR code~\cite{li2022unleashing}. For instance, Cummins et al.~\cite{cummins2023large} proposed a 7B LCM to optimize LLVM assembly code, while 01.AI~\cite{mlmproject} developed a ``machine language model'' to analyze executable programs and binary code. 

\subsection{Multi-Lingual Inconsistency in Large Language Models}

The performance gap of LLMs between high-resourced languages and underrepresented languages has been studied recently in various natural language processing tasks \cite{Lai2023ChatGPTBE, Bang2023AMM, Shi2022LanguageMA, Wang2023AllLM, Peng2023TowardsMT}, such as logical reasoning~\cite{Foroutan2023BreakingTL}, nature language understanding~\cite{Huang2023NotAL}, and nature language generation~\cite{Deng2023MultilingualJC}. Specifically, \cite{Shi2022LanguageMA} revealed that GPT-3 \cite{Brown2020LanguageMA} and PaLM \cite{anil2023palm} perform worse when answering the math questions in underrepresented languages, compared with the performance of answering the same questions in English. The work \cite{Wang2023AllLM} found that LLMs produce significantly more unsafe responses for non-English queries than English ones, indicating the unsatisfied safety alignment for non-English languages. 
These performance gaps can be attributed to the language-imbalanced nature of training data and alignment data: The majority of training and alignment data of current LLMs, such as GPT-4, are in English~~\cite{Bang2023AMM, alves2024tower, zhao2024llama, Achiam2023GPT4TR}. Meanwhile, previous works have proposed several methods to transfer knowledge from high-resource languages to underrepresented languages. 
Additionally, the work \cite{Foroutan2023BreakingTL} proposes an attention mechanism that uses a dedicated set of parameters to encourage cross-lingual attention in code-switched sequences. 
Another thread of work adopts prompting strategies. \cite{Huang2023NotAL} design a generic template prompt that stimulates cross-lingual and logical reasoning skills to enhance task performance across languages. 
In addition, \cite{Deng2023MultilingualJC} utilizes the LLM to generate multilingual training data, which is then used for fine-tuning the LLM, to alleviate the performance gap without any human labeling effort. 


Different from previous works, this paper focuses on code generation tasks, which have yet to be explored, and investigates the two aspects of multi-lingual bias, including multi-NL understanding and multi-PL generation. Existing work in the code intelligence field only reports the performance in different PLs \cite{wang2022no, wang2023prompt, gao2023two}, ignoring the bias among PLs. Contrarily, our work conducts extensive experiments to investigate how to improve the code generation performance and multi-lingual bias of LCMs.

\section{Conclusion}\label{sec:con}
In this paper, we have identified the multi-NL and multi-PL bias of LCMs in the code generation task. In addition, we have explored to mitigate the bias in the prompting and instruction tuning phase. Extensive experiments demonstrate that both prompting and instruction tuning aid in mitigating the multi-lingual bias, and instruction tuning can further boost the code generation capabilities of LCMs. Moreover, we provide insights and implications for researchers and developers aimed at mitigating the multi-lingual bias and improving the performance of LCMs.

\bibliographystyle{IEEEtran}
\bibliography{cite,bib/zj,bib/ref}

\end{document}